\documentclass[12pt]{article}
\usepackage{amsmath}
\usepackage{amssymb}
\usepackage{epsfig}
\usepackage{euscript}
\def\mathswitchr#1{\relax\ifmmode{\mathrm{#1}}\else$\mathrm{#1}$\fi}

\newcommand {\pslash}{\hbox{$\not\hbox{\kern-2.3pt $p$}$}}

\usepackage[compress]{cite}
\usepackage{epic}

\def\alf1{ {\alpha\over\pi} }

\begin{document}
\begin{titlepage}
\title{Einstein-Heisenberg Consistency Condition Interplay with Cosmological Constant Prediction in Resummed Quantum Gravity }


\author{B.F.L. Ward\\
    Baylor University\\
        Waco, TX, USA\\
        bfl\_ward@baylor.edu}

\maketitle

\begin{abstract}
We argue that our recent success in using our resummed quantum gravity approach to Einstein's general theory of relativity, in the context of the Planck scale cosmology formulation of Bonanno and Reuter, to estimate the value of the cosmological constant $\Lambda$ supports the use of quantum mechanical consistency requirements to constrain the main uncertainty in that very promising result. This main uncertainty, which is due to the uncertainty in the value of the time $t_{\text{tr}}$ at which the transition from the Planck scale cosmology to the FRW model occurs, is shown to be reduced, by requiring consistency between the Heisenberg uncertainty principle and the known properties of the solutions of Einstein's equations, from four orders of magnitude to the level of a factor of ${\cal O}(10)$. This lends more credibility to the over-all resummed quantum gravity approach itself, in general, and to our estimate of $\Lambda$ in particular.
\end{abstract}
\centerline{                BU-HEPP-15-02, ~May,~ 2015}
\end{titlepage}

In Ref.~\cite{darkuni}, we have used the theory of resummed quantum 
gravity~\cite{bw1,bw2,bw2a,bw2b,bw2c,bw2d,bw2e,bw2f,bw2g,bw2h,bw2i} to estimate the value of the cosmological constant as 
$\rho_\Lambda=(0.0024 eV)^4 $ which is quite close to the experimental 
value~\cite{cosm11,cosm12,pdg2008} $\rho_\Lambda|_{\text{expt}}\cong ((2.37\pm 0.05)\times 10^{-3}eV)^4 $.
As we have emphasized in the Ref.~\cite{darkuni}, the main source of uncertainty in our formula for $\rho_\Lambda$ in Eq.(53) in the latter reference is the value of the time 
$t_{\text{tr}}$ at which the universe switches from the Planck scale 
cosmology~\cite{reuter1,reuter2} of Bonanno and Reuter to the FRW model that we see today.
Following Refs.~\cite{reuter1,reuter2}, we took the value $t_{\text{tr}}\cong 25/M_{\text{Pl}}$ as suggested by numerical studies. We have argued that this value could be uncertain by a couple of orders of magnitude, so that our estimate for $\rho_\Lambda$ could be off by as much as four orders of magnitude. In what follows, we use the union of the Heisenberg uncertainty principle and the known properties of the solutions~\cite{nactmn,duerr,cherni-tag} of the Einstein equations for de Sitter space to arrive at a consistency condition which reduces the uncertainty in $t_{\text{tr}}$  to the level of ${\cal O}(3)$. This adds significantly to the credibility of our original estimate.

Specifically, using the theory of resummed quantum gravity in the context of the
Planck scale cosmology of Refs.~\cite{reuter1,reuter2}, we have arrived at the estimate
\begin{equation}
\begin{split}
\rho_\Lambda(t_0)&\cong \frac{-M_{Pl}^4(1+c_{2,eff}k_{tr}^2/(360\pi M_{Pl}^2))^2}{64}\sum_j\frac{(-1)^{F_j}n_j}{\rho_j^2}\cr
          &\qquad\quad \times \frac{t_{tr}^2}{t_{eq}^2} \times (\frac{t_{eq}^{2/3}}{t_0^{2/3}})^3\cr
    &\cong \frac{-M_{Pl}^2(1.0362)^2(-9.194\times 10^{-3})}{64}\frac{(25)^2}{t_0^2}\cr
   &\cong (2.4\times 10^{-3}eV)^4,
\end{split}
\label{eq-rho-expt}
\end{equation}
where we take the age of the universe to be $t_0\cong 13.7\times 10^9$ yrs.  
Here, the connection between $\rho_\Lambda$ and the cosmological 
constant $\Lambda$ is as usual: $\rho_\Lambda=\frac{\Lambda}{8\pi G_N}$, 
where $G_N$ is Newton's constant. The constant $c_{2,eff}$ is estimated in Ref.~\cite{darkuni} as $2.56\times 10^4$ from the known elementary particles in the universe. The sum over $j$ in (\ref{eq-rho-expt}) is over all elementary particles where $F_j,\; n_j$ are their respective fermion number and effective number of degrees of freedom. The quantity $\rho_j$ is given 
by $\ln\frac{2}{\lambda_c(j)}$
where $\lambda_c(j)=\frac{2m_j^2}{\pi M_{Pl}^2}$ when $m_j$ is the rest mass of particle $j$. Here $M_{Pl}$ is the Planck mass. The time $t_{eq}$ is the time of matter-radiation equality and we have followed here the arguments of Ref.~\cite{branch-zap1,branch-zap2}. We also note that Refs.~\cite{sola1,sola2} have made similar 
arguments for the connection between cosmological time and the effective running scales in (\ref{eq-rho-expt}) and have made analysis that leads to a qualitatively similar result for $\rho_\Lambda(t_0)$. What is interesting is the closeness of the result in (\ref{eq-rho-expt}) to the experimental value~\cite{cosm11,cosm12,pdg2008}:
$\rho_\Lambda(t_0)|_{\text{expt}}\cong ((2.37\pm 0.05)\times 10^{-3}eV)^4.$
This suggests that the uncertainty on our estimate for $t_{\text tr}$, which we have put~\cite{darkuni} at the level of a factor of 100, is in fact much less than this latter value. In what follows, we argue that this is indeed the case.

The basic physical idea which we wish to apply here is the known property of a de Sitter universe, which we describe here with the metric~\cite{rn,ratra} 
$$g_{\mu\nu}dx^{\mu}dx^{\nu}=dt^2-e^{2t/b}[dw^2+w^2(d\theta^2+\sin^2\theta d\phi^2)]$$ in an obvious notation, with $b=\sqrt{3/\Lambda}$
: if a light ray starts at the origin ($w=0$ here) and travels uninterruptedly, it never gets past the point $w=w_0\equiv b$ along its geodesic. If we treat the quantum mechanics as truly interwoven with the fabric of space-time, as it most certainly should be, according to Einstein's general theory of relativity, it must know about the latter limit for the quantum wave function of the photons in this light ray. According to the Heisenberg uncertainty principle, the uncertainty associated with the momentum conjugate variable to the coordinate distance $w$ is correspondingly bounded in the quantum theory of general relativity. To get a realization of the attendant constraint, we use the results in Refs.~\cite{nactmn,duerr,cherni-tag,ratra} to 
check for the consistency of this bound with the effective scale $k$ associated to the running values of $G_N(k),\; \Lambda(k)$ as implied by the methods of Refs.~\cite{reuter1,reuter2,sola1,wein1,reutera,laut,reuterb,reuter3,litim1,litim2,perc1,perc2,perc3,perc4,perc5} and by our resummed quantum gravity (RQG) 
approach.

Specifically, we use the basic formulation of the Heisenberg uncertainty principle,
\begin{equation}
\Delta{p}\Delta{q}\ge \frac{1}{2},
\label{eheq1}
\end{equation}
where we define $\Delta{A}$ as the quantum mechanical uncertainty of the observable $A$ and $p$ is the momentum conjugate to the observable coordinate $q$. In our case, we have $q=w\cos\theta$ where $\theta$ is the polar angle when the direction of $\vec{k}$ is taken along the $\hat{z}$ direction and we may identify $\Delta{p}$ as our effective $k$, as $k$ represents the size of the mean squared momentum fluctuations in the universe that are effective for the running of the universe observables  $G_N(k),\; \Lambda(k)$. As we see from the explicit solutions of the field equations in Refs.~\cite{nactmn,duerr,cherni-tag}, for the universe in the Planck regime, the solutions of the scalar field equations,\footnote{Spin continues to be an inessential complication here~\cite{mlgbgr}.} in an appropriate set of coordinates, are spanned by plane waves in 3-space with Bessel/Hankel function-related dependence on time, so that we have the estimate, at any given time, again using an obvious notation, 
\begin{equation}
(\Delta{q})^2\cong \frac{\int_{0}^{w_0}dw w^2 w^2<\cos^2\theta>}{\int_{0}^{w_0}dw w^2}=\frac{1}{5}w_0^2.
\label{eheq2}
\end{equation} 
From this estimate, we get the Einstein-Heisenberg consistency condition
\begin{equation}
k\ge \frac{\sqrt{5}}{2w_0}=\frac{\sqrt{5}}{2}\frac{1}{\sqrt{3/\Lambda(k)}}
\label{eheq3}
\end{equation}
where $\Lambda(k)$ is given by Eq.(52) in Ref.~\cite{darkuni}:
\begin{equation}
\begin{split}
\Lambda(k)
         &=\frac{-\pi M_{Pl}^2(k)}{8}\sum_j\frac{(-1)^{F_j}n_j}{\rho_j^2}\cr
         &=\frac{-\pi M_{Pl}^2(1+c_{2,eff}k^2/(360\pi M_{Pl}^2))}{8}\sum_j\frac{(-1)^{F_j}n_j}{\rho_j^2}\cr
         &\cong \frac{\pi M_{Pl}^2(1+c_{2,eff}k^2/(360\pi M_{Pl}^2))\times 9.194\times 10^{-3}}{8}
\end{split}
\label{eheq4}
\end{equation}
Indeed, when $k$ becomes too small to satisfy the condition (\ref{eheq3}), we argue that the Planck scale inflation must end. Thus, we have the estimate of the transition time, $t_{\text tr}=\alpha/M_{Pl}=1/k_{\text{tr}}$, from the Planck scale inflationary regime to the Friedmann-Robertson-Walker regime via the value of $\alpha$ for which equality holds in (\ref{eheq3}). On our solving for $\alpha$ we get 
\begin{equation}
\alpha\cong 25.3,
\label{eheq5}
\end{equation}
which is in a agreement with the value $\alpha\cong 25$ implied by the numerical studies in Ref.~\cite{reuter1,reuter2}.

In conclusion, we argue that we have significantly reduced the theoretical uncertainty of the estimate of $\Lambda$~\cite{darkuni} from the standpoint of the application of our RQG theory in the context of the Planck scale cosmology of Refs.~\cite{reuter1,reuter2}.
In particular, using the union of the ideas of Heisenberg and Einstein, 
we argue that the previous estimate of an uncertainty of a factor of ${\cal O}(100)$ on $t_{\text tr}$ is now reduced to the level of a factor of ${\cal O}(3)$, so that the uncertainty of our estimate on $\Lambda$ is now reduced to a factor of
${\cal O}(10)$. In our view, this represents considerable progress in the long campaign to understand the value
of $\Lambda$ from {\it first principles} using the method of the operator-valued field.

\section*{Acknowledgments}
We thank Profs. L. Alvarez-Gaume and W. Hollik for the support and kind
hospitality of the CERN TH Division and the Werner-Heisenberg-Institut, MPI, Munich, respectively, where a part of this work was done.
\par


\begin{thebibliography}{00}
\bibitem{darkuni} B.F.L. Ward, Phys. Dark Universe {\bf 2} (2013) 97.
\bibitem{bw1} B.F.L. Ward, Open Nucl.Part.Phys.Jour. {\bf 2}(2009) 1.
\bibitem{bw2} B.F.L. Ward, Mod. Phys. Lett. A{\bf 17} (2002) 2371.
\bibitem{bw2a} B.F.L. Ward, Mod. Phys. Lett. A{\bf 19} (2004) 143.
\bibitem{bw2b} B.F.L. Ward, J. Cos. Astropart. Phys.{\bf 0402} (2004) 011.
\bibitem{bw2c} B.F.L. Ward, hep-ph/0605054, Acta Phys. Polon. {\bf B37} (2006)
1967.
\bibitem{bw2d} B.F.L. Ward, hep-ph/0503189, Acta Phys. Polon. {\bf B37} (2006) 
347.
\bibitem{bw2e} B.F.L. Ward, hep-ph/0502104, in {\it Focus on Black Hole Research}, ed. P.V. Kreitler,(Nova Sci. Publ., Inc., New York, 2006) p. 95.
\bibitem{bw2f} B.F.L. Ward, hep-ph/0411050, Int. J. Mod. Phys. {\bf A20} (2005) 3502.
\bibitem{bw2g} B.F.L. Ward, hep-ph/0411049, Int. J. Mod. Phys. {\bf A20} (2005)
3128. 
\bibitem{bw2h} B.F.L. Ward, hep-ph/0410273, in {\it Proc. ICHEP 2004}, {\bf vol. 1}, eds. H. Chen {\it et al.},(World Sci. Publ. Co., Singapore, 2005) p. 419 and references therein.
\bibitem{bw2i} B.F.L. Ward, Mod. Phys. Lett. A{\bf 23} (2008) 3299.
\bibitem{cosm11} A.G. Riess {\it et al.}, Astron. Jour. {\bf 116} (1998) 1009.
\bibitem{cosm12} S. Perlmutter {\it et al.}, Astrophys. J. {\bf 517} (1999) 565;
and, references therein.
\bibitem{pdg2008} C. Amsler {\it et al.}, Phys. Lett. B{\bf 667} (2008) 1.
\bibitem{reuter1} A. Bonanno and M. Reuter, Phys. Rev. D{\bf 65} (2002) 043508.
\bibitem{reuter2} A. Bonanno and M. Reuter, Jour. Phys. Conf. Ser. {\bf 140} (2008) 012008, arXiv:0803.2546; and references therein.
\bibitem{nactmn} O. Nachtmann, Commun. Math. Phys. {\bf 6} (1967) 1.
\bibitem{duerr} G. Boerner and H.P. Duerr, Nuovo Cimento {\bf 64} (1969) 669, and references therein.
\bibitem{cherni-tag} N.A. Chernikov and E.A. Tagirov, Ann. Inst. H. Poincare {\bf 9} (1968) 109.
\bibitem{branch-zap1} V. Branchina and D. Zappala, G. R. Gravit. {\bf 42} (2010) 141. 
\bibitem{branch-zap2} V. Branchina and D. Zappala, arXiv:1005.3657, and references therein.
\bibitem{sola1} I. L. Shapiro and J. Sola, Phys. Lett. B{\bf 475} (2000) 236.
\bibitem{sola2} J. Sola, J. Phys. A{\bf 41} (2008) 164066.
\bibitem{rn} H.P. Robertson and T.W. Noonan,{\it Relativity and Cosmology},(Saunders, Philadelphia, 1968).
\bibitem{ratra} B. Ratra, Phys. Rev. D{\bf 31} (1985) 1931, and references therein. 
\bibitem{wein1} S. Weinberg, in {\it General Relativity, an Einstein Centenary Survey},
eds. S. W. Hawking and W. Israel, (Cambridge Univ. Press, Cambridge, 1979).
\bibitem{reutera} M. Reuter, Phys. Rev. D{\bf 57} (1998) 971, and references therein.
\bibitem{laut} O. Lauscher and M. Reuter, Phys. Rev. D{\bf 66} (2002) 025026, and references
therein.
\bibitem{reuterb} E. Manrique, M. Reuter and F. Saueressig, Ann. Phys. {\bf 326} (2011) 440 and references therein.
\bibitem{reuter3}
A. Bonanno and M. Reuter, Phys. Rev. D{\bf 62} (2000) 043008, and references therein.
\bibitem{litim1}
D. F. Litim, Phys. Rev. Lett.{\bf 92}(2004) 201301; Phys. Rev. D{\bf 64} (2001)
105007.
\bibitem{litim2} P. Fischer and D.F. Litim, Phys. Lett. B{\bf 638} (2006) 497 and references therein.
\bibitem{perc1} D. Dou and R. Percacci, Class. Quant. Grav. {\bf 15} (1998) 3449.
\bibitem{perc2} R. Percacci and D. Perini, Phys. Rev. D{\bf 67}(2003) 081503.
\bibitem{perc3} R. Percacci and D. Perini, Phys. Rev. D{\bf 68} (2003) 044018.
\bibitem{perc4} R. Percacci, Phys. Rev. D{\bf 73}(2006) 041501.
\bibitem{perc5} A. Codello, R. Percacci and C. Rahmede, Int.J. Mod. Phys. A{\bf 23}(2008) 143, and references therein.
\bibitem{mlgbgr} M.L. Goldberger, private communication.
\end{thebibliography}
\end{document}